# Do Rogue Wave Exist in the Kadomtesv-Petviashivili I Equation ?

Jie-Fang Zhang $^{a,b}$, Zhao Zhang $^{c}$, Meng-yang Zhang $^{d}$, and Mei-zhen Jin $^{e}$

a) Institute of Intelligent Media Technology, Communication University of Zhejiang, Hangzhou 310018, Zhejiang, China;

b) Zhejiang Provincial Key Laboratory of Film and Television Media Technology, Communication University of Zhejiang, Hangzhou 310018, Zhejiang, China;

c) Guangdong Provincial Key Laboratory of Nanophotonic Functional Materials and Devices, South China Normal University, Guangzhou 510631, China;

d) Covestro (Shanghai) Investment Company Limited, Shanghai 201507, China;

e) Network and Data Center, Communication University of Zhejiang, Hangzhou 310018, Zhejiang, China.

**Abstract:** There is considerable fundamental theoretical and applicative interest in obtaining two-dimensional rogue wave similar to one-dimensional rogue wave of the nonlinear Schrödinger equation. Here, we first time proposes a self-mapping transformation and analytically predict the existence of a family of novel spatio-temporal rogue wave solutions for the Kadomtesv-Petviashivili equation. We discover that these spatio-temporal rogue waves showing a strong analogy characteristics of the 'short-lives' with rogue waves of the NLS equation. Our fingdings can also provide a solid mathematical basis for theory and application in shallow water, plasma and optics. This technique could be available to construct rogue-like waves of (2+1)-dimensional nonlinear wave models. Also, these studies could be helpful to deepen our understandings and enrich our knowledge about rogue waves.

# 1. Introduction

As an extreme natural phenomenon, rogue waves have attracted intense investigation from both experimentalists and theoreticians on different perspective because of its importance in their concerned fields [1-18]. Theoretically, rational solutions of nonlinear Schrödinger (NLS) equation play a major role in the study of rogue waves for the mathematical description.

It is well known that the NLS equation is only a wave model suitable for open ocean or deep water areas, but rogue waves frequently hit beaches and coastal areas, accounting for nearly 70 percent of the total number of extreme water wave events[19]. Moreover, the mathematical description of water waves in shallow water is usually based on solutions of the Korteweg-de Vries (KdV) equation or Kadomtsev-Petviashvili (KP) equation [20,21], where the NLS equation are not valid. Although a great deal of effort has been made to explain extreme waves in the open seas [1,2,4], little has been done to describe the rogue waves that occur in coastal areas and shallow waters [22,23].

It is also known that the KdV equation exist *N*-soliton solutions [24] and and the KP equation exist *N*-line soliton solutions[25], respectively. Especially the KPI equation possesses also the space localized traveling lump solutions[26]. Though the lump is not exactly similar to the spatio-temporal localized Peregine soliton of the NLS equation, they are all expressed rational functions and decay algebraically. Therefore, a very natural question whether exist the exact pure algebraic and spatio-temporal localized solution in the KdV type equation or the KP type equation? This question is worth pondering, because, as far as we know, the spatio-temporal localized solutions of two-dimensional nonlinear evolution models, that is, the rogue wave solutions, has not really been found.

Self-similarity is a fundamental physical property that has been extensively studied in many areas of physics[27-29]. This is because the existence of a self-similar solution for a given nonlinear evolution equation is of great importance in understanding widely different nonlinear physical phenomena [30]. It was shown that linearly chirped parabolic pulses are approximate self-similar solutions of NLS equation with gain in the high-intensity limit [31,32]. These results have been confirmed experimentally [33-35] and have extended previous theoretical studies of parabolic pulse propagation in optical fibers[36,37]. Kruglov et al. [38] first introduced an interesting self-similar technique to search for self-similar solutions to the generalized NLS equation with distributed dispersion, nonlinearity, and gain or loss. Ponnomarenko et al.[39] had extended this analysis method to show that bright and dark spatial self-similar waves can propagate in graded-index amplifiers exhibiting the self-focusing or the self-defocusing Kerr nonlinearities. Moreover, Belmonte-Beitia et al.[40] had constructed also explicit nontrivial solutions of NLS equations with potentials and nonlinearities depending both on time and on the spatial coordinates. Senthilnathan et al.[41] utilized also the self-similar analysis to the case of cubic-quintic optical media, where the chirped bright soliton solutions have been obtained in the anomalous and normal dispersion regimes. Recently, Triki et al. [42] investigated further the propagation characteristics of the chirped self-similar solitary waves in non-Kerr nonlinear media within the framework of the generalized NLS equation with distributed dispersion, two power-law nonlinearitis, and gain or loss. Liang et al. [43] advance a statistical theory of extreme event emergence in random nonlinear wave systems with self-similar intermediate asymptotics and show that extreme events and even rogue waves in weakly nonlinear, statistical open systems emerge as parabolic-shape giant fluctuations in the self-similar asymptotic propagation regime. It should be noted that the research literature by using of self-similar analysis method theoretically and observing to self-similarity experimentally are very rich after more than 20 years of development, which can not be listed due to space limited, here. Until now, no attempts had been made to find a self-mapping transformation(SMT) for a specific nonlinear wave models, including the NLS equation itself. We think that this is not only different with self-similar transformation(SST) from one model to another model, but also this problem is also of prime importance.

During present study, we noticed that the similarity reductions of the KP equation by Lie group method[44,45] and by the direct reduction method[46] are thoroughly studied and obtained all reduction of the KP equation to ordinary differential equations and to the (1+1)-dimensional partial differential equations [47-50]. Lou et al. [48,49] found more arbitrary functions included in these results which have been missed by the classical Lie approach. Clarkson et al. [50] gave the result of applying the direct method to obtain one-step reductions of the KP equation to ordinary differential equations. But, to the best of our knowledge, the investigation about the self-similar transformation in the KP equations similar to those of the NLS equation, are still open problems.

Our research is motivated by the following reasons: (1) Does the self-similarity as the NLS equation exists in the KP equation? (2) Can one find these new spatio-temporal localized solutions for the KP equation which is similar to the similaritons and rogue waves in NLS equation? In this letter, we will give a positive answer the above the question by taking the KPI equation as an example for using the self-similarity.

## 2. SMT of the KPI equation

We consider the KPI equation in the following form

$$[u_t + 6uu_x + u_{xxx}]_x - u_{yy} = 0, \qquad (1)$$

where $u = u(x, y, t)$ and $x, y$ and $t$ are space variables and time variable, respectively, (subscripts denote partial derivatives). Our goal is to research for a self-mapping transformation for KPI equation itself, that is to become the KPI equation(1) into the KPI equation in following from

$$(U_\tau + 6UU_\xi + U_{\xi\xi\xi})_\xi - U_{\eta\eta} = 0, \qquad (2)$$

where $U = U(\xi, \eta, \tau)$, $\xi, \eta$ and $\tau$ are new introducing space variables and time variable, respectively. To connect solutions of (1) with those of (2), we introduce the following self-mapping transformation(SMT) for KPI equation (1)

$$u(x, y, t) = \rho U(\xi, \eta, \tau) + H, \qquad (3)$$

where $U(\xi, \eta, \tau)$ satisfies KP equation (2); $\xi = \xi(x, y, t), \eta = \eta(y, t), \tau = \tau(t)$ are three relational expressions between dependent variables $(x, y, t)$ in Eq.(1) and dependent variables $(\xi, \eta, \tau)$ in Eq.(2); $H = H(x, y, t) = f(t)x + g(t)y + h(t)y^2 + e(t)$ is a excitation background; $\rho = \rho(t)$ is a amplitude scaling factor. $\xi(x, y, t), \eta(y, t)$, $\tau(t), \rho(t), f(t), g(t), h(t), e(t)$ are all functions of the indicated variables, which may be obtained by substituting SMT(3) into Eq.(1). Requiring $U(\xi, \eta, t)$ to satisfy the KP equation (2) and $u(x, y, t)$ to be a solution of the KP equation (1), one get the set of equations

$$f_t + 6f^2 - 2h = 0, \tag{4}$$

$$\xi_{xx} = 0, \eta_{yy} = 0, \tag{5}$$

$$\xi_x \xi_t - \xi_y^2 + 6\xi_x^2(fx + gy + hy^2 + e) = 0, \tag{6}$$

$$\rho_t \xi_x + \rho \xi_{xt} - \rho \xi_{yy} + 12\rho f \xi_x = 0, \tag{7}$$

$$\xi_x \eta_t - 2\xi_y \eta_y = 0, \tag{8}$$

$$\tau_t = \rho \xi_x = \xi_x^3 = \frac{\eta_y^2}{\xi_x}. \tag{9}$$

It can be inferred from (5) and (7)

$$\xi(x, t) = \kappa(t)x + \iota(t)y^2 + \gamma(t)y + \omega(t), \eta(y, t) = m(t)y + n(t), \tag{10}$$

where $\kappa(t), \iota(t), \omega(t), m(t), n(t)$ are five undetermined functions of the specified time variable $t$. Inserting (10) into (5)–(7) and some tedious algebraic calculations yields

$$\rho = \kappa^2, m = \kappa^2, \tau = \int_0^t \kappa^3 dt,$$

$$f = -\frac{\kappa_t}{6\kappa}, \iota = \frac{\kappa_t}{2}, \gamma = \frac{n_t}{2\kappa}, \tag{11}$$

$$h = \frac{2\kappa_t^2 - \kappa \kappa_{tt}}{12\kappa^2}, g = \frac{3n_t \kappa_t - \kappa n_{tt}}{12\kappa^3}, e = \frac{n_t^2 - 4\kappa^3 \omega_t}{24\kappa^4}.$$

It is interesting to note that when we have a known solution of the KP equation (2), a novel solution the KPI equation (1) can be obtained as the results of the SMT of the solution.

It should be especially mentioned here that we got the reference [51] when we basically completed present study. In this reference, Nishitani and Tajiry obtained an invariant transformation of the KPII equation by means of finite transformation and studied the resonant interactions of similar-type solitons moving in non-steady and non-uniform background. It is verified that our results can be used to restore the results given in [51], but our findings are applicable to both KPI and KPII equations. Moreover, our research method is direct and simple, which can be extended to other nonlinear evolution models.

In view of the importance of self-similarity, we especially discuss the self-mapping transformation of the KPI equation with self-similarity as follows:

$$u(x,y,t) = \kappa^2 U(\xi,\eta,\tau) - \frac{\kappa_t}{6\kappa}x + \frac{3n_t\kappa_t - n_{tt}\kappa}{12\kappa^3}y + \frac{n_t^2 - 4\kappa^3\omega_t}{24\kappa^4}, \qquad (12)$$

with

$$\xi = \kappa x + \frac{\kappa_t}{2}y^2 + \frac{n_t}{2\kappa}y + \omega(t), \quad \eta = \kappa^2 y + n(t), \quad \tau = \int_0^t \kappa^3 dt, \qquad (13)$$

where $\kappa = \dfrac{\kappa_0}{1+6f_0 t}$. The amplitude of the self-similar solution mentioned above is directly proportional to $(6f_0 t + 1)^{-2}$, and it possesses a non-steady and non-uniform background. Specifically, when $\omega(t) = 0$ and $n(t) = 0$, the parameter $\kappa_0$ controls the initial amplitude, while the parameter $f_0$ in $f(t) = \dfrac{f_0}{1+6f_0 t}$ governs the degree of background inclination and the decaying rate of solution.

As can be seen above, although this SMT of the KP equation, of which formulation is written in terms of self-similarity involving some parameter without subject to any constraints, is similar to the widely used SST, what we consider is a SMT of the same nonlinear evolution model instead of a SST between two nonlinear evolution models. As far as we know, this has not been reported in the literature. The SMT of KPII equation can be similarly obtained. Because KPII equation doesn't exist the space localized lump solutions, only the corresponding results for KPI equation are given here.

### 3. Spatio-temporal rogue-waves of the KPI equation

As application of the SMT (12) with (13), in what follow, we present the novel decaying self-similar solutions existed in KP equation (1).

As we know that the KP equation (2) has a two-dimensional solitary wave called a lump[52]:

$$U(\xi,\eta,\tau) = 4a^2 \frac{3a^4\eta^2 - a^2(3a^2\tau - \xi)^2 + 1}{\left[a^2(3a^2\tau - \xi)^2 + 3a^4\eta^2 + 1\right]^2}, \tag{14}$$

where $a$ is an arbitrary real parameter. The lump wave described by the aforementioned equation reaches its maximum value $4a^2$ at coordinate $(3a^2t, 0)$, and it exhibits identical minimum values $-a^2/2$ at coordinates $(3a^2t + \sqrt{3}/a, 0)$ and $(3a^2t - \sqrt{3}/a, 0)$.

Based on SMT (12) with (13), another novel decaying quickly self-similar lump of the KPI equation moving on a nonuniform background $H(x,y,t)$ can be obtained. In this paper, we only consider the special case of $\omega(t) = 0$ and $n(t) = 0$. In this scenario, the self-similar solution corresponding to Eq. (14) exhibits a more concise form:

$$u(x,y,t) = \frac{4a^2\kappa_0^2}{(6f_0t+1)^2} \frac{-a^2\theta^2 + 3a^4\kappa_0^4 y^2/(6f_0t+1)^4 + 1}{a^2\theta^2 + 3a^4\kappa_0^4 y^2/(6f_0t+1)^4 + 1} + \frac{f_0 x}{6f_0t+1}, \tag{15}$$

where $\theta = \frac{3a^2\kappa_0^3 t(3f_0t+1)}{(6f_0t+1)^2} - \frac{\kappa_0 x}{6f_0t+1} + \frac{3\kappa_0 f_0 y^2}{(6f_0t+1)^2}$. The solution described by the above equation can be interpreted as a motion that accelerates slowly and tends to a constant speed on an inclined plane, with its amplitude proportional to $(6f_0t+1)^{-2}$. This wave reaches its maximum value $u_{max} = \frac{a^2\kappa_0^2(9t^2f_0^2 + 3f_0t + 4)}{(6f_0t+1)^2}$ at coordinate

$$\left\{x = \frac{3\kappa_0^2 a^2 t}{2} + \frac{a^2\kappa_0^2}{4f_0} - \frac{a^2\kappa_0^2}{4f_0(6f_0t+1)}, y = 0\right\}, \tag{16}$$

with a relative amplitude $A = \dfrac{4a^2\kappa_0^2}{(6f_0 t+1)^2}$. As time progresses, the relative amplitude tends to zero, and at this point, the background value at the wave peak is $\dfrac{a^2\kappa_0^2}{4}$. More precisely, the relationship between the background value at the wave peak and the relative amplitude can be described as follows: $H(A) = \dfrac{a^2\kappa_0^2}{4} - \dfrac{A}{16}$. This implies that while the parameter $f_0$ can affect the decay time and the initial slope of the background, regardless of its selection, the maximum value decays to one-sixteenth of its initial value ($t=0$).

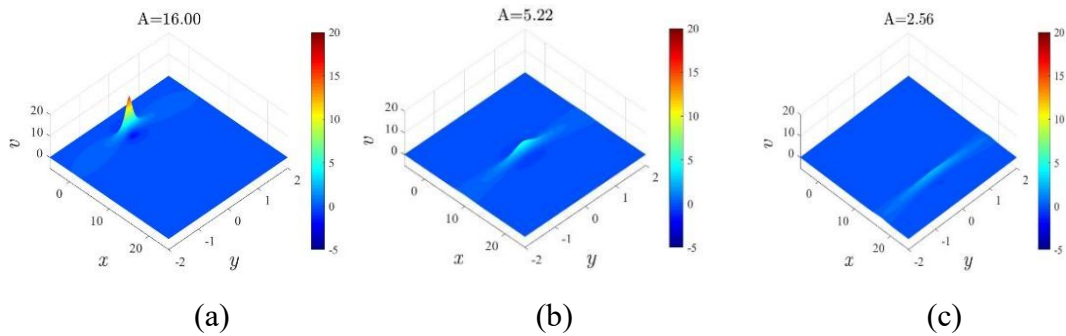

(a)　　　　　　　　　　(b)　　　　　　　　　　(c)

Fig.1. (Color online.) Time evolution of the self-similar rogue wave of KPI equation. Here $v(x,y,t) = u(x,y,t) - H(x,t)$. The parameters are chosen as $f_0 = 1/8$, $\kappa_0 = 2, a = 1$. (a) t=0 ; (b) t=1; (c) t=2.

　　　As can be seen from Figure 1 that this decaying quickly self-similar lump has the characteristic similar to rogue wave described by the Peregine soliton of the NLS equation except that it moves with time $t$. Furthermore, it can be observed from Figure 1 that the spatial locality of the lump solution weakens as time progresses. That is, due to the dissipation effect, the wave width of the lump wave becomes wider with the passage of time. Therefore, the lump wave molecule [53] of Eq. (2) is transformed into unsteady structures by SMT (12) with Eq.(13). For instance, through the utilization of the Hirota transformation:

$$U(\xi,\eta,\tau) = 2\dfrac{\partial^2}{\partial \xi^2} \ln F(\xi,\eta,\tau), \qquad (17)$$

the simplest molecule of the KPI equation can be expressed as

$$F = 14\left[V^2\eta^2 + V(V\tau-\xi)^2\right]^3 + \left[238V^2\eta^2 + 70V(V\tau-\xi)^2\right]\left[V^2\eta^2 + 5V(V\tau-\xi)^2\right]$$
$$- 4bV^{3/2}(V\tau-\xi)^3 + 12bV^{5/2}(V\tau-\xi)\eta^2 - 1750V(V\tau-\xi)^2 + 6650V^2\eta^2 \qquad (18)$$
$$+ 4b\sqrt{V}(V\tau-\xi) + 26250 + 2b^2/7.$$

The lump wave molecule described in the above equation consists of three lump waves, with a horizontal velocity of $V$ and a vertical velocity of 0. The stable structure described by Eq. (18) undergoes SMT, resulting in dissipative phenomena that cause a slow increase in the distance between these three lump waves over time. However, the decay rate of these three lump waves with respect to time is identical, proportional to $(6f_0 t + 1)^{-2}$, as shown in Figure 2. And comparing Figure 1 and Figure 2, it can be found that the first-order lump wave and the second-order lump wave have the same attenuation rate.

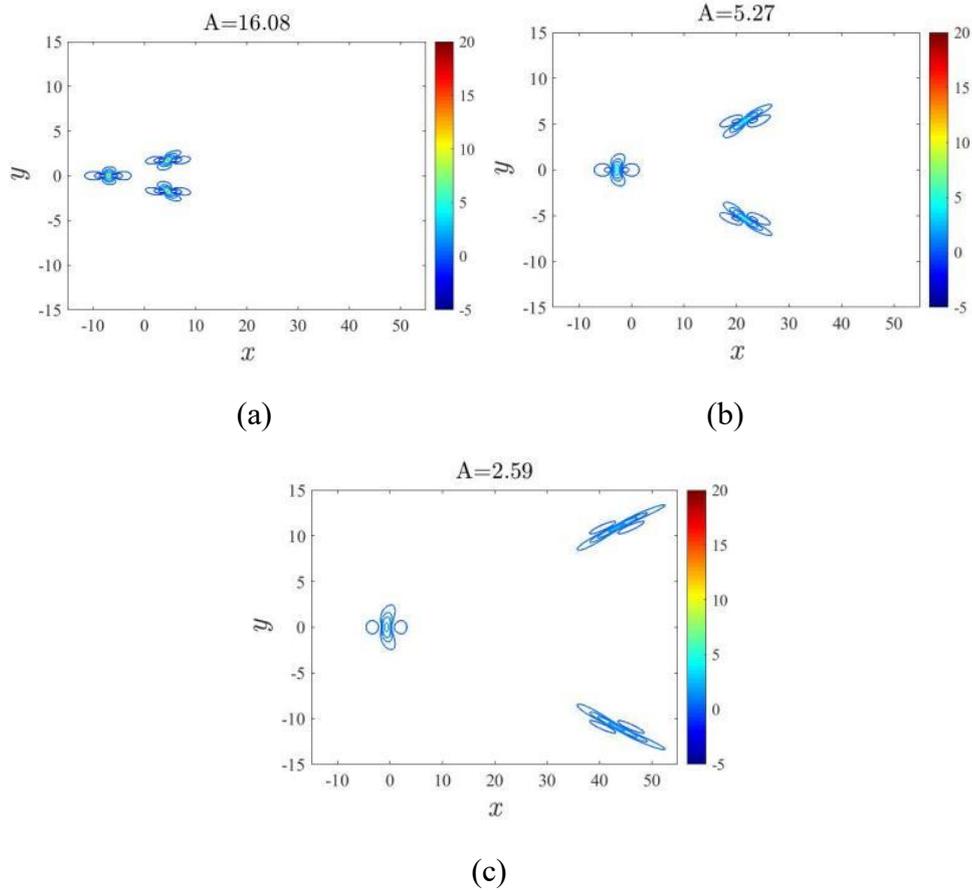

Fig.2. (Color online.) The stable structure described by Eq. (18) undergoes SMT to become an unstable structure, where the relevant parameters are $f_0 = 1/8, \kappa_0 = 2, a = 1, b = 10^5$. Presented herein is the relative amplitude A. (a) $t = 0$; (b) $t = 1$; (c) $t = 2$.

## 4. Conclusion

To summarize, we have established a new SMT of the KPI equation itself and discovered a class of novel rogue waves of the KPI equation. Our results show also definitively that the amplitude of the wave controlled by various parameters can be large and is decayed very quickly in a short time, so that it just describes the characteristics of rogue waves.

The presented results could not directly be generalized to the other (2+1)-dimensional nonlinear evolution models, but it can bring some enlightenment for studying various integrable and non-integrable nonlinear models by using of the SMT or the SST. The significance of our findings is not restricted to water rogue waves, which can be applied to nonlinear optics and to other fields where the KP equation can be used the governing model. We surmise that new experiments characterized by KP equation will be easier to implement and the predictions of the present work could maybe be verified.

Furthermore, SMT(3) or (12) with Eq.(13) also imply that partial differential equations with constant coefficients can be transformed into bilinear equations [54] with variable coefficients. This paradigm shift has the potential to disrupt conventional solution procedures and consequently unveil a plethora of novel exact solutions within integrable systems.

### Acknowledgments

Our contributions to this field have benefited from valuable collaborations and discussions with numerous colleagues and friends, in particular Prof. Da-Jun Zhang and Jing-song He as well as Doctor Lei Wu. Jie-fang Zhang extends further thanks to Prof. Wen-xiu Ma and Prof. Zhi-jun Qiao for sparking his interest in the wider aspects of the self-similarity when he visited the United States.

This work is supported by the NSF of China under Grant No.61877053.

### References

[1] C. Kharif, E. Pelinovsky, and A. Slunyaev, Rogue Waves in the Ocean,

*Springer-Verlag, Berlin* (2009).

[2] M. Onorato, S. Residori, U. Bortolozzo, A. Montina, and F. Arecchi, Rogue waves and their generating mechanisms in different physical contexts, *Phys. Rep.* **528** (2013) 47–89.

[3] N. S. Ginzburg, R. M. Rozental, A. S. Sergeev, A. E. Fedotov, I. V. Zotova, and V. P. Tarakanov, Generation of Rogue Waves in Gyrotrons Operating in the Regime of Developed Turbulence, *Phys. Rev. Lett.* **119** (2017) 034801.

[4] T. A. A. Adcock, and P. H. Taylor, The physics of anomalous ('rogue') ocean waves, *Rep. Prog. Phys.* **77** (2014) 105901.

[5] Y. Y. Tsai, J. Y. Tsai, and I. Lin, Generation of acoustic rogue waves in dusty plasmas through three-dimensional particle focusing by distorted waveforms, *Nature Phys.* **12(6)** (2016) 573-577

[6] P. S. Vinayagam, R. Radha, K. Porsezian, Taming rogue waves in vector Bose–Einstein condensates, *Phys. Rev. E* **88** (2013) 042906.

[7] J. M. Soto-Crespo, N. Devine, and N. Akhmediev, Integrable turbulence and rogue waves: breathers or solitons? *Phys. Rev. Lett.* **116** (2016) 103901.

[8] N. S. Ginzburg, R. M. Rozental, A. S. Sergeev, A. E. Fedotov, I. V. Zotova, and V. P. Tarakanov, Generation of rogue waves in gyrotrons operating in the regime of developed turbulence, *Phys. Rev. Lett.* **119** (2017) 034801.

[9] N. Akhmediev, J. M. Dudley, D. R. Solli, and S. K. Turitsyn, Recent progress in investigating optical rogue waves, *J. Opt.* **15** (2013) 060201.

[10] J. M. Dudley, F. Dias, M. Erkintalo, and G. Genty, Instabilities, breathers and rogue waves in optics, *Nature Photo.* **8** (2014) 755–764.

[11] K. Hammani, B. Kibler, C. Finot, P. Morin, and G. Millot, Peregrine soliton generation and breakup in standard telecommunications fiber, *Opt. Lett.* **36** (2011) 112–114.

[12] M Shats, H. Punzmann, and H. Xia, Capillary rogue waves, *Phys. Rev. Lett.* **104** (2010) 104503.

[13] J. F. Han, T. Liang, and W. S. Duan, Possibility of the existence of the rogue wave and the super rogue wave in granular matter, *Euro. Phys. J. E* **42** (2019) 5.

[14] M. G. Copus, and R. E. Camley, Creation of magnetic rogue waves, *Phys. Rev. B* **102** (2020) 220410(R).

[15] D. R. Solli, C. Ropers, P. Koonath, and B. Jalali, Optical rogue waves, *Nature* **450** (2007) 1054-1057.

[16] A. Chabchoub, N. P. Hoffmannn, and N. Akhmediev, Rogue wave observation in a water wave tank, *Phys. Rev. Lett.* **106** (2011) 204502.

[17] H Bailung, S. K. Sharma, Y. Nakamura, Observations of peregrine solitons in a multi- component plasma with negative ions, *Phys. Rev. Lett.* **107** (2011) 255005.

[18] H. Xiong, J. H. Gan, and Y. Wu, Kuznetsov-Ma soliton dynamics based on the mechanical effect of light, *Phys. Rev. Lett.* **119** (2017) 153901.

[19] I. Nikolkina and I. Didenkulova, Rogue waves in 2006-2011, *Nat. Hazards Earth Syst. Sci.* **11** (2011) 2913.

[20] T. Soomere, Rogue waves in shallow water, Eur. *Phys. J. Spec. Top.* **185** (2010) 81-96.

[21] Y. Kodama, KP solitons in shallow water, *J. Phys. A: Math. Theor.* **43** (2010) 434004.

[22] H. I. Abdel-Gawad, M. Tantawy, and R. E. A. Elkhair, On the extension of solutions of the real to complex KdV equation and a mechanism for the construction of rogue waves, *Waves Random Complex Media* **26** (2016) 397.

[23] A. Ankiewicz, M. Bokaeeyan, and N. Akhmediev, Shallow-water rogue waves: An approach based on complex solutions of the Korteweg-de Vries equation, *Phys.Rev.E* **99** (2019) 050201(R).

[24] R. Hirota, Exact solution of the Korteweg-de Vries equation for multiple collisions of solitons, *Phys. Rev. Lett.* **27** (1971) 1192.

[25] J Satsuma, N-soliton solution of the two-dimensional Korteweg-de Vries equation, J. Phys. Soc. Jpn. **40(1)** (1976) 286-290.

[26] S. V. Manakov, V. E. Zakharov, L. A. Bordag, V. B. Matveev, Two-dimensional solitons of the Kadomtsev-Petviashvili equation and their interaction, *Phys. Lett. A* **63**(1977) 205-206.

[27] G. I. Barenblatt, Scaling, self-similarity, and intermediate asymptotics (*Cambridge University Press, Cambridge, England,* (1996).

[28] J. M. Dudley, C. Finot, D. J. Richardson, and G. Millot, Self-similarity in ultrafast nonlinear optics, *Nat. Phys.* **3** (2007) 597.

[29] N. S. Ginzburg, R. M. Rozental, A. S. Sergeev, A. E. Fedotov, I. V. Zotova, and V. P. Tarakanov, Generation of rogue waves in gyrotrons operating in the regime of developed turbulence, *Phys. Rev. Lett.* **119** (2017) 034801 .

[30] P. L. Sachdev, Self-similarity and beyond: exact solutions of nonlinear problems, *CRC, New York,* (2000).

[31] M. E. Fermann, V. I. Kruglov, B. C. Thomsen, J. M. Dudley, and J. D. Harvey, Self-similar propagation and amplification of parabolic pulses in optical fibers, *Phys. Rev. Lett.* **84** (2000) 6010.

[32] V. I. Kruglov, A. C. Peacock, J. M. Dudley, and J. D. Harvey, Self-similar propagation of high-power parabolic pulses in optical fiber amplifiers, *Opt. Lett.* **25**(2000) 1753.

[33] V. I. Kruglov, A. C. Peacock, J. D. Harvey, and J. M. Dudley, Self-similar propagation of parabolic pulses in normal-dispersion fiber amplifiers, *J. Opt. Soc. Am. B* **19** (2002) 461.

[34] K. D. Moll, and A. L. Gaeta, and G. Fibich, Self-similar optical wave collapse: observation of the townes profile, *Phys. Rev. Lett.* **90** (2003) 203902.

[35] C. Finot, G. Millot, C. Billet, and J. M. Dudley, Experimental generation of parabolic pulses via Raman amplification in optical fiber, *Opt.Express* **11** (2003) 1547.

[36] S. Boscolo, S. K. Turitsyn, V. Yu. Novokshenov, and J. H. B. Nijhof, On the theory of self-similar parabolic optical solitary waves, *Theor. and Math. Phys.* **133** (2002) 1647.

[37] F. Ö. Ilday, J. R. Buckley, W. G. Clark, and F. W. Wise, Self-similar evolution of parabolic pulses in a laser, *Phys. Rev. Lett.* **92** (2004) 213902.

[38] V. I. Kruglov, A. C. Peacock, and J. D. Harvey, Exact self-similar solutions of the generalized nonlinear Schrödinger equation with distributed coefficients, *Phys. Rev. Lett.* **90** (2003) 113902.

[39] S. A. Ponomarenko and G. P. Agrawal, Do solitonlike self-similar waves exist in nonlinear optical media, *Phys. Rev. Lett.* **97** (2006) 013901.

[40] J. Belmonte-Beitia, V. M. Pérez-García, V. Vekslerchik, and V. V. Konotop, Localized nonlinear waves in systems with time- and space-modulated nonlinearities, *Phys. Rev. Lett.* **100** (2008) 164102.

[41] A. Choudhuri, H. Triki, and K. Porsezian, Self-similar localized pulses for the nonlinear Schrdinger equation with distributed cubic-quintic nonlinearity, *Phys. Rev. A* **94** (2016) 063814.

[42] H. Triki, K. Porsezian, K. Senthilnathan, and K. Nithyanandan, Chirped self-similar solitary waves for the generalized nonlinear Schrödinger equation with distributed two-power-law nonlinearities, *Phys. Rev. E* **100** (2019) 042208.

[43] C.-H Liang, S. A. Ponomarenko, F. Wang, and Y.-J Cai, Rogue waves, self-similar statistics, and self-similar intermediate asymptotics, *Phys. Rev. A* **100** (2019) 063804.

[44] G.W. Bluman and J.D. Cole, Similarity Methods for Differential Equations, Applied Mathematical Science No.**13** *Springer, Berlin* (1974).

[45] P. J. Oiver, Applications of Lie Groups to Differential Equations, Graduate Texts in Mathematics No. 107 (Springer, New York, 1986).

[46] P.A. Clarkson and M.D. Kruskal, New Similarity Reductions of the Boussinesq Equation, *J. Math. Phys.* **30** (1989) 2201-2213.

[47] M. Tajiri, T. Nishitani and S. Kawamoto, Similarity Solutions of the Kadomtsev-Petviashvili Equation, *J. Phys. Soc. Japan* **51** (1982) 2350-2356.

[48] S.-Y. Lou, Phys. Similarity solutions of the Kadomtsev-Petviashvili equation, *J. Phys. A: Math. Gen.* 23 (1990) L649-L654.

[49] S.-Y. Lou, H-Y Ruan, D-F Chen, and W.-Z Chen, Similarity reductions of the KP equation by a direct method, *J. Phys. A: Math. Gen.* **24** (1991) 1455-1467.

[50] P. A. Clarkson and and P. Winternitz, Nonclassical symmetry reductions for the Kadomtsev- Petviashvili equation, *Physica D* **49** (1991) 257.

[51] T. Nishitani and M. Tajiri, Invariant Transformation of the Kadomtsev-Petviashvili Equation, *J. Phys. Soc. Jpn.* **53** (1984) 79.

[52] J. Satsuma, M.J. Ablowitz, Two-dimensional lumps in nonlinear dispersive systems, J. Math. Phys. **20 (7)** (1979) 1496-1503.

[53] Z. Zhang, X. Yang, B. Li, Q. Guo, and Y. Stepanyants, Multi-lump formations from lump chains and plane solitons in the KPI equation, *Nonlinear Dynamics*, **111(2)** (2023) 1625-1642.

[54] R. Hirota, The direct method in soliton theory. Cambridge University Press, (2004).